\documentclass{aa} 
\usepackage{graphicx}
\begin{document}
  \authorrunning{Guenther et al.}
  \titlerunning{BS Indi: An enigmatic object}

   \title{BS Indi : An enigmatic object in the Tucana association$^*$}


         \author{E.W. Guenther\inst{1}
          \and
          E. Covino\inst{2}
          \and
          J.M Alcal\'a\inst{2}
          \and
          M. Esposito\inst{3}
          \and
          R. Mundt\inst{4}
             }

   \offprints{Eike Guenther, \email{guenther@tls-tautenburg.de}\\$*$~
    Based on observations obtained at the European Southern
    Observatory at La Silla, Chile in program 
    62.I-0418, 63.I-0096, 64.I-0294, 65.I-0012, 
    67.C-0155, 68.C-0292, 69.C-0207, 70.C-0163, 
    072.A-9012, 073.C-0355 and 67.C-0194}

   \institute{Th\"uringer Landessternwarte Tautenburg,
              Sternwarte 5, D-07778 Tautenburg, Germany
          \and
              INAF-Osservatorio Astronomico di Capodimonte
              via Moiariello, 16 -
              I 80131 Napoli,
              Italy
          \and
              Universit\`a di Salerno, via S. Allende - 
              I 84081 Baronissi (SA) -
              Italy 
          \and
              MPI f\"ur Astronomie, K\"onigstuhl 17, 
              D - 69117 Heidelberg, Germany
	      }

   \date{Received; accepted}

   \abstract{BS Ind (=HD 202947) is a young K0V star in the Tucana
    association. Photometric observations with the Hipparcos satellite
    show an eclipse-like light-curve with primary and secondary
    eclipse. The eclipsing binary has a period of 0.435338 days and a
    circular orbit. Our spectroscopic observations however show that the
    K0V primary is a single-line spectroscopic binary with a period of
    3.3 years. The minimum mass of the invisible component is about
    0.9~$M\odot$ which means that the mass of the companion is about the
    same as that of the primary. The first inspection of our FEROS
    spectra with a resolution of 48000, as well as a CES spectrum with a
    resolution of 220000 shows no obvious companion.  However, when the
    FEROS spectra are cross-correlated with an M-star, a secondary
    becomes visible as a broad peak in the cross-correlation
    function. The width and the position of this broad peak is variable
    on a short time.  When phased to a period of 0.435338 days,
    the radial velocity variations of the broad peak show the
    characteristic sine-wave of a spectroscopic binary in a circular
    orbit. The best interpretation of this data is that the broad peak
    in the cross-correlation function is caused by an eclipsing binary
    consisting of two late-K, or early-M stars with an orbital period of
    0.435338 days. This is the eclipsing system. These two stars then
    orbit the K0V-primary with a period of 3.3 years. The assumption
    that BS Ind is a triple system consisting of a K0V star and
    two late-K, or early-M stars also explains the unusual brightness of
    the object and the near infrared excess. Thus, BS Ind is
    unique, as it contains by far the shortest-period young binary star,
    and these stars are eclipsing.

    } \maketitle


\section{Introduction}

In recent years a number of loose associations of young nearby stars
have been identified.  Among these new groups is the Tucana association
(Torres et al.\cite{Torres00}) which has an age of about 30 Myr and is
located at a distance of about 45 to 60 pc. The Tucana association might
be related to the close-by Horologium association (Zuckerman \& Webb
2000; Zuckerman, Song, \& Webb 2001) which has about the same distance
and age. In total, about 50 stars have been identified in these
associations. Because of the small distance, these associations are well
suited for all kinds of studies of young stars.  In the first step we
survey these stars for the presence of visual companions (Neuh\"auser et
al. \cite{neuhauser03}). Similar to our survey for spectroscopic
binaries of the TW~Hydrae association (Torres et al. \cite{torres03}),
we have also surveyed the stars in the Horologium and Tucana
associations.

\object{BS Ind} (\object{HIP 105404}, \object{HD 202947}) is one
of the stars in the Tucana association. It is classified as a K0V star
located at a distance of $46\pm3$~pc (Hipparcos, ESA \cite{ESA97}).
With an absolute brightness of $M_v=5.59$ mag (Zuckerman et
al. \cite{zuckerman01}), or $M_v=5.54$ mag (Cutispoto et
al. \cite{cutispoto02}), \object{BS Ind} is thus slightly above
the main sequence. From our spectra, we derive an equivalent width of
$0.176\pm0.020$~\AA \, for the Li\,I\,6707 line, which also confirms the
young age of \object{BS Ind}. In summary, at first glance
\object{BS Ind} seems to be a fairly normal low-mass star with an age
of about 30 Myr located in the Tucana association.


\section{The light-curve of BS Ind}

\begin{figure}[h]
\resizebox{\hsize}{!}{\includegraphics{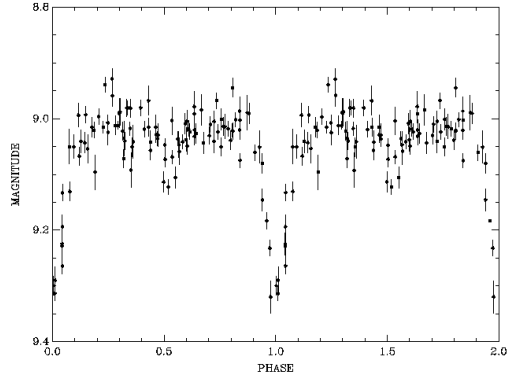}}
\caption{Hipparcos light-curve phase-folded with the period from the
light-curve of 0.435338 days.  The primary eclipse is clearly visible,
and the secondary eclipse is barely visible.  }
\label{light-curve}
\end{figure}

The Hipparcos catalogue lists \object{BS Ind} as an eclipsing binary
with a period of 0.435338 days (Hipparcos, ESA \cite{ESA97}).  Is it
really an eclipsing binary?  The phase-folded light-curve of 
\object{BS Ind} in fact looks like that of a typical eclipsing binary, with
primary and secondary eclipses (Fig.\,\ref{light-curve}).  Additionally,
the secondary eclipse appears, though barely above the noise, at phase
0.5, as is expected for a companion on a circular orbit.  The depth
of the primary eclipse is $0.26\pm0.01$ mag, and that of the secondary
$0.10\pm0.01$ mag in the Hipparcos V-band.  Although the data is quite
noisy, we estimate the time difference between the first and the fourth
contact as $1.5\pm0.2$ hours, and the time difference between the second
and the third contact as $0.5\pm0.2$ hours. A spectrum taken by
Cutispoto et al. (\cite{cutispoto02}), shows a double-line emission
core in a Ca~II K-line but not in the LiI 6707-line. The implication of
this observation will be discussed in section~4.

In principle stellar spots can mimic an eclipsing binary light-curve
(Joergens et al. \cite{joergens01}).  The question thus arises whether
we can exclude this possibility. A spot should be visible
half of the time, if the star is viewed close to equator-on, and even
longer if it is viewed more pole-on.  This is certainly not what we
observe. Secondly, from our spectra we derive a $v\sin{i}$ of
$13.0\pm0.4$ km\,s$^{-1}$ which is in excellent agreement with the value
from the literature (Zuckerman \& Webb \cite{zuckerman00}; Cutispoto et
al. \cite{cutispoto02}).  If the light-curve were caused by spots, the
true rotation period would have to be 0.435338 days, which implies that
the inclination should be less than 10 degrees. If the inclination would
be that low, it is hard to imagine how spots could produce an 0.3 mag
dip in the light-curve. The spot would have to cover more than 20\% of
the stellar surface and it would need to be hidden most of the time,
thus it would have to be located close to the equator.  However, a spot
located close to the equator on a star that is viewed almost pole-on can
not produce a rapid decline like the one observed
(Fig.\,\ref{light-curve}).

There is further evidence against the hypothesis of such a short
rotation period: like other stars of the same age, \object{BS Ind} is
an X-ray source.  The logarithmic X-ray luminosity varies between 29.4
and 30.4 $\log({erg\,s^{-1}})$. This value can be compared to the $log$ of the
mean X-ray luminosity of $28.94\pm0.06$ $\log({erg\,s^{-1}})$ for K-stars in the
Pleiades which are older than \object{BS Ind}, and $29.78\pm0.10$
$\log{erg\,s^{-1}}$ for WTTS in the Taurus star-forming region which are
younger than \object{BS Ind} (Stelzer \& Neuh\"auser
\cite{stelzer00}).  The X-ray brightness of \object{BS Ind} thus is
what is expected for a K-star of this age. Furthermore, the $log$ of
X-ray luminosity is $29.7$ $\log({erg\,s^{-1}})$ for stars with a rotation
period of 4 days in the Taurus star-forming region and $29.2$
$\log({erg\,s^{-1}})$ for stars with a rotation period of 4 days in the
Pleiades, respectively. Thus, the X-ray brightness of \object{BS Ind}
is typical for a K-star of the same age that has a rotation period of
about 4 days.  A rapidly rotating star should be much brighter at
X-rays.  The hardness ratios are also within the range of other stars in
Tucana (Kastner et al. \cite{kastner03}). Thus, the X-ray brightness and
the $v\sin{i}$ both support the idea that \object{BS Ind} is an
eclipsing system and not a very rapidly rotating star with huge,
unevenly distributed, photospheric spots.

\section{The long-period orbit}

In order to solve the orbit of \object{BS Ind}, we observed
the star with the ESO echelle spectrograph FEROS (Fiber-fed Extended
Range Optical Spectrograph) on the 1.5-m, and 2.2-m-telescope at La
Silla.  The spectra cover the wavelength region between about 3600 \AA
\ and 9200 \AA, and have a resolution of $\rm \lambda / \Delta
\lambda=48000$.  The standard MIDAS pipeline for FEROS was used for bias
subtraction, flat-fielding, scattered light removal, echelle order
definition and extraction, and wavelength calibration of the spectra.
Precise measurements of the radial-velocity (RV) have then been carried
out by cross-correlating the object spectra with spectra of the RV
standard \object{HR 5777} observed with the same instrument using the
IRAF RV package.  The RV of \object{HR 5777} is known to be stable, and
has been determined with an accuracy of $\rm 55\,ms^{-1}$ (Murdoch
\cite{murdoch93}). We have observed \object{HR 5777} in every observing
run and find that the scatter of our measurements is $\rm 80\,ms^{-1}$,
which is consistent with the expected errors of the measurements.  The
cross-correlation function between \object{BS Ind} and \object{HR
5777} shows only one peak.  Because of the smaller signal to noise ratio
and the larger $v\,\sin{i}$, the errors on the RV-measurements of
\object{BS Ind} are often larger (see Table\,\ref{tab:rv}).  We
also took one spectrum of \object{BS Ind} with the CES - Coude
Echelle Spectrometer at the 3.6-m-telescope at La Silla in the
wavelength region between 6402 - 6445 \AA \, with the High-resolution
image slicer.  This instrument has a resolution,
$\lambda/\Delta\lambda$, of 220000. The spectrum of \object{{\bf BS
Ind}} was exposed for twice 1200s. Standard IRAF routines were used for
the data analysis.


\begin{table}
\caption{Radial velocity measurements and corresponding rms errors
versus Heliocentric Julian Day.
The first column indicates the running number of the spectrum. 
($*$ taken from Cutispoto et al. (2002))}
\begin{tabular}{llr}
\hline \hline
No. & HJD & RV [km\,s$^{-1}$] \\
\hline
Cutispoto $*$ & 2449253.5 & $31.0\pm1.1$ \\
Cutispoto $*$ & 2449597.5 &  $9.0\pm1.8$ \\
FEROS No. 1 & 2451737.9193 & $32.53\pm0.04$ \\
FEROS No. 2 & 2452031.9386 & $ 8.54\pm0.10$ \\
FEROS No. 3 & 2452089.8273 & $ 5.88\pm0.20$ \\
FEROS No. 4 & 2452093.8171 & $ 6.74\pm0.05$ \\
FEROS No. 5 & 2452093.8513 & $ 6.72\pm0.05$ \\
FEROS No. 6 & 2452093.8823 & $ 6.75\pm0.06$ \\
FEROS No. 7 & 2452093.9113 & $ 7.14\pm0.09$ \\
FEROS No. 8 & 2452093.9404 & $ 6.95\pm0.11$ \\
FEROS No. 9 & 2452093.9532 & $ 6.79\pm0.08$ \\
CES         & 2452105.9038 & $ 6.91\pm0.16$ \\
FEROS No.10 & 2452372.9120 & $ 5.76\pm0.10$ \\
FEROS No.11 & 2452372.9224 & $ 5.62\pm0.05$ \\
FEROS No.12 & 2452373.8833 & $ 5.50\pm0.05$ \\
FEROS No.13 & 2452373.8961 & $ 5.22\pm0.08$ \\
FEROS No.14 & 2452373.9073 & $ 4.86\pm0.09$ \\
FEROS No.15 & 2452384.9085 & $ 5.66\pm0.07$ \\
FEROS No.16 & 2452385.9122 & $ 5.71\pm0.13$ \\
FEROS No.17 & 2452710.8900 & $ 7.87\pm0.14$ \\
FEROS No.18 & 2452717.9110 & $ 8.42\pm0.05$ \\
FEROS No.19 & 2452723.9049 & $ 8.65\pm0.06$ \\
FEROS No.20 & 2453096.9020 & $13.55\pm0.11$ \\
FEROS No.21 & 2453104.87233 & $12.74\pm0.09$ \\
FEROS No.22 & 2453105.89532 & $12.97\pm0.05$ \\
FEROS No.23 & 2453137.88899 & $11.03\pm0.05$ \\
FEROS No.24 & 2453137.89652 & $10.99\pm0.05$ \\
FEROS No.25 & 2453137.90404 & $10.24\pm0.14$ \\
FEROS No.26 & 2453137.90475 & $10.32\pm0.05$ \\
FEROS No.27 & 2453162.87508 & $ 9.56\pm0.10$ \\
FEROS No.28 & 2453162.88265 & $ 9.59\pm0.06$ \\
FEROS No.29 & 2453168.78650 & $10.34\pm0.06$ \\
FEROS No.30 & 2453168.79417 & $10.22\pm0.06$ \\
FEROS No.31 & 2453168.83513 & $ 9.76\pm0.07$ \\
FEROS No.32 & 2453168.84271 & $ 9.69\pm0.06$ \\
\hline\hline
\end{tabular}
\label{tab:rv}
\end{table}

Table\,\ref{tab:rv} gives the RV measurements obtained for \object{BS
Ind}. An eclipsing M0V-star would have a mass of about $0.5M_\odot$
which will induce RV-variations with an amplitude of more than 200
km\,s$^{-1}$, if it is orbiting the primary star with a period of
0.435338 days, and using $M_1= 0.8 M_\odot$ for the primary. The much
smaller amplitude observed already indicates that either we observed the
object incidentally always at the same phase, or the mass of the
companion is much lower, or the orbital period is much longer.
Fig.\,\ref{BS_Indi-phaseIII} shows the RV measurements phase-folded with
the period 0.435338 days. The dashed line is for
$f(m)={{M_2^3\sin^3{i}}\over{(M_1+M_2)^2}}=0.07~M_\odot$, which
corresponds to $K_1=116$ km\,s$^{-1}$. As a systemic velocity we use the
average RV of stars in the Tucana association of $20.6\pm4.6$, and we
set $\sin{i}=1$, because the system is supposed to be
eclipsing. Clearly, this curve does not fit the data. Note that
error-bars of our RV-measurements are so small that they cannot be seen
in the plot. Even if we assume a much lower mass, the RV-measurements
still cannot be reproduced with this period. \object{BS Ind} clearly is
not a simple eclipsing binary.

\begin{figure}[h]
\resizebox{\hsize}{!}{\includegraphics{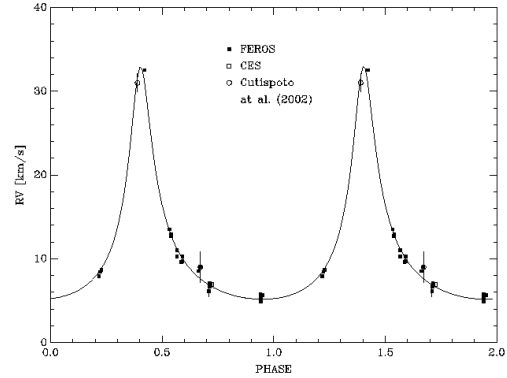}}
\caption{The most likely orbital period is 3.3 years.}
\label{orbit}
\end{figure}

As can clearly be seen from Table\,\ref{tab:rv}, \object{BS Ind} does
show significant RV-variations. However, on {\bf HJD 2452093} we took 6
spectra within 3.3 hours. The scatter of 0.15 km\,s$^{-1}$ of these
data-points is fully consistent with the accuracy of the measurements,
and already excludes a very short period, massive companion.  Spectra
taken from HJD 2452710 to HJD 2452723 differ also only by $0.78\pm0.15$
km\,s$^{-1}$, and spectra taken between HJD 2453096 and HJD 2453105 only
by $0.85\pm0.15$ km\,s$^{-1}$. This indicates that a period of the
radial velocity-variations below 10 days is not possible. If we add in
the two data-points from Cutispoto et al. (\cite{cutispoto02}), we find
a period of 3.3 years, and a mass function $f(m)={{M_2^3
sin^3i}\over{(m_1+m_2)^2}}$ of $0.273\pm0.06$ $M\odot$ which implies a
minimum mass of the companion of $0.9\pm0.1$ $M\odot$ assuming a mass of
the primary of $0.8\pm0.1$ $M\odot$.  The resulting fit is shown in
Fig.\,\ref{orbit} and the orbital parameters derived are given in
Tab.\,\ref{tab:orbit}. The systemic velocity, $R_0=10.8\pm0.2$
km\,s$^{-1}$ is notably different from the average RV of stars in the
Tucana association of $20.6\pm4.6$ km\,s$^{-1}$.

\begin{table}
\caption{Orbit of BS Ind}
\begin{tabular}{ll}
\hline \hline
period                                    & $1223\pm30$ days \\ 
eccentricity                              & $0.60\pm0.05$ \\    
$\omega$                                  & $-6\pm2^o$  \\
$R_0$                                     & $10.8\pm0.2$ km\,s$^{-1}$ \\
$f(m)={{M_2^3 sin^3i}\over{(m_1+m_2)^2}}$ & $0.273\pm0.06\,M_\odot$ \\ 
$a\,sin\,i$                               & $2.7\pm0.2$ AU    \\
$m_1$                                     & $0.8\pm0.1\,M_\odot$ \\
$m_2$                                     & $0.9\pm0.1\,M_\odot$ \\
periastron passage [JD]                   & $2450489\pm30$ \\    
\hline\hline
\end{tabular}
\label{tab:orbit}
\end{table}

\section{The detection of the two eclipsing stars}

The best explanation for the fact that the K0V-primary is a long-period,
single-line, spectroscopic binary, and the light-curve shows a short
period eclipsing system is that \object{BS Ind} is a triple
system. 
The eclipsing components are two late-K, or early-M stars orbiting each
other with a period of 0.435338 days, and this binary system orbits the
K0V-primary with a period of 3.3 years. In order to prove this scenario
the two eclipsing stars have to be detected in the spectrum.




As we pointed out before, when we cross-correlate the FEROS spectra with
the K1IV-star \object{HR 5777} there is only one peak.
Fig.\,\ref{BS_Indi-specCaIIK} shows three of our FEROS spectra of this
region. In none of these spectra we see any obvious companion. On the
other hand, the spectral region around the Ca~II K-line is very crowded
and thus might not be the best region to look for a faint companion.
As mentioned above, a spectrum taken by Cutispoto et
al. (\cite{cutispoto02}) show a double-line structure in the Ca~II
K-line. Because this feature was observed only once, it is apparently
highly variable. Since the Ca~II -lines are prominent during
flare-events it is possible that the second peak was caused by a
flare-event. This idea is also consistent with the absence of a
double-line peak in the Li\,I\,6707 line.  The wavelength region
between 6402 - 6445 \AA \, contains a number of unblended and well
separated spectral lines and is thus well-suited for the search of the
secondary component.  Fig.\,\ref{BS_Indi_CES} shows a CES spectrum with
a resolution of $\lambda/\Delta\lambda$$=220000$ of the FeI 6430.856
line of the K5V star \object{$\rho$ Eri} together with spectrum of
\object{BS Ind}. Again there is no obvious sign of the companions.
The strongest feature in the spectrum of an early M-star are the
TiO-band-heads.  Fig.\,\ref{TiOspectrum} shows the spectrum of
\object{BS Ind} together with a properly scaled M star. Even the
averaged spectrum of \object{BS Ind} does not allow to exclude, or
confirm the presence of such a companion.


\begin{figure}[h]
\resizebox{\hsize}{!}{\includegraphics{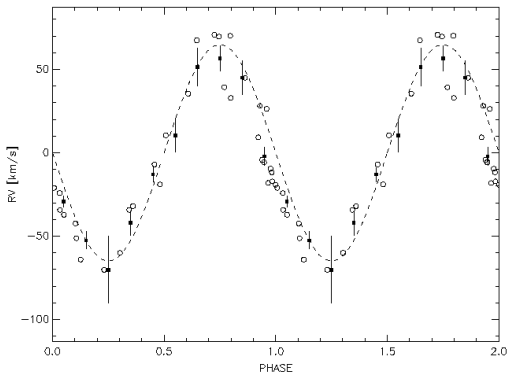}}
\caption{The RV curve derived for the broad peak, after
subtracting the 3.3 year orbit, and after phase-folding the data to the
period of 0.435338 days.  The filled dots are the phase averaged data
points, and the open circles are the individual measurements.  The phase
is the same as in Fig.\,\ref{light-curve}. The eclipse in fact appears,
at RV=0, and before the primary eclipse the spectrum is red-shifted, just
as it should be.  Thus, there is clear evidence that the position of the
broad peak varies with a period of 0.435338 days. The semi-amplitude is
about 65 km\,s$^{-1}$.}
\label{BS_Indi-phaseII}
\end{figure}

As a last step we finally cross-correlated the FEROS spectra of
\object{BS Ind} with the M0III star \object{HR 6056}.  In this case we
did detect a broad peak (Fig.\,\ref{BS_Indi_min_max_ccM}).  This broad
peak is observed in all our FEROS spectra when cross-correlated with the
M-star. The position and the width of this peak is variable, and we
regard it as the signature of the two eclipsing early-M, or late-K K
stars.  After subtracting the 3.3 years orbit from this data, and
phase-folding it with the 0.435338 day period, we find in fact a
sinusoidal variation of the RV (Fig.\,\ref{BS_Indi-phaseII}).  The phase
in this figure is the same as in Fig.\,\ref{light-curve}. The eclipse
thus in fact appears, at RV=0, and before the primary eclipse the
spectrum is red-shifted, just as it should be for an eclipsing binary.
The two eclipsing stars appear only as one broad peak, because the
$v\,\sin{i}$ of stars are of the order of 100 km\,s$^{-1}$. This is
because the orbital period and the rotational period of the two stars is
the same, because the orbital period is so short. Because of the large
$v\,\sin{i}$ the lines of the two stars are not separated, and because
one is brighter than the other, the width and the position of the broad
peaks in the cross-correlation function vary.  This is fully consistent
with the light-curve, which also indicates that one of the eclipsing
stars is brighter. Unfortunately this implies that amplitude of the
radial-velocity variations of the broad peak is only the change of the
barycenter of the blended cross-correlation functions of the two
stars. The true amplitude of the RV-variations is larger.  This implies
that although the B and C component of \object{{\bf BS\,Ind}} are
eclipsing, we cannot determine their true masses.  We thus conclude that
\object{BS Ind} in fact consist of two eclipsing stars orbiting each
other with a period of 0.435338 days, and this system orbits the
K0V-primary with a period of 3.3 years.

\section{Conclusions}

The photometry from the Hipparcos satellite indicates the presence of
two eclipsing stars orbiting each other with a period of 0.435338
days. {\object{BS Ind} thus has a lot in common with the multiple star
\object{HD 97770}, which also contains two eclipsing late type stars
(K4/5 V star and a K5V star) , with an almost similar orbital period of
$0.476533\pm0.000033$ days (Cutispoto et al. \cite{cutispoto97}). Like
\object{BS Ind}, \object{HD 97770} shows flares and is bright in
X-rays. The main difference is that \object{BS Ind} is possibly younger
than \object{HD 97770}}. The light-curve of \object{BS Ind} cannot be
explained by a spot on the star. The RV measurements clearly indicate
that \object{{\bf BS Ind}} is not a short-period eclipsing object but an
SB1-binary with an orbital period of 3.3 years. The mass of the
secondary component is, however, about one solar mass. The
cross-correlation function does show, apart from the sharp peak due to
the K0V star, a very broad peak which varies significantly in RV and
width. This can best be explained as the signature of two stars orbiting
each other.  The two eclipsing stars, whose total mass is about 0.9
$M\odot$, are thus either late-K or early-M stars. Both the light-curve
and the cross-correlation function indicate that the brightness of the
two eclipsing stars is slightly different. The presence of the two
eclipsing stars also explains the total brightness of the \object{BS
Ind}-system in the optical and in the near infrared regime. Assuming
that the two eclipsing stars have about 0.5 $M\odot$, they would be
roughly M0V stars and thus together would be 2.1 mag fainter than the
primary in the visible. If this were true, we would expect a
decrease in brightness by 0.15 mag if one of the stars is eclipsed. This
value is is in the middle between the primary and secondary eclipse. In
reality one of the eclipsing stars thus must be a slightly brighter and
the other a slightly fainter. The system can best be disentangled once
we know the true brightness of the primary and the eclipsing system
separately, which will be possible once VLTI measurements are obtained.
If we subtract the flux of two M0V-stars from the measured absolute
brightness of the whole \object{{\bf BS Ind}}-system, we find that the
K0V-primary has an absolute brightness of $M_v=5.7\pm0.1$ and would thus
be essentially on the main-sequence, like the other stars in Tucana.  If
the system would consist of two M0V stars and a K0V star it should be
7.2, 6.6, and 6.5 mag in the J, H, and K-band, respectively instead of
7.5, 7.0 and 7.0 for single K0V-star. This is again in perfect agreement
with the 2MASS point source catalogue which gives values of
$7.184\pm0.026$, $6.699\pm0.031$, and $6.574\pm0.024$ mag for J, H, and
K-band for \object{BS Ind}, respectively.

Thus, the fact that \object{{\bf BS Ind}} is a triple system consisting
of a K0V star and two stars with a spectral type of about M0V not only
explains our spectroscopic observations and the Hipparcos light-curve,
but also the optical and near infrared fluxes.  Thus, \object{BS Ind} is
by far the shortest period young low-mass binary system
known. Unfortunately, the radial-velocities of the two eclipsing stars
cannot be measured individually, and thus the masses of these stars
cannot be measured individually. However, the fact that there is a young
binary with an orbital period of only 0.435338 days, raises the question
of how such a system could have formed. The possible observation of
a flare by Cutispoto et al. \cite{cutispoto02} also fits nicely into the
picture that the K0-primary is orbited by two late-type stars of very
short period. Such binaries are active and thus show flare-activity.
 
Addtionally, the separation between the primary and the
two eclipsing stars is large enough to be resolved with VLTI.
In this way it will be possible to obtain the true mass of the
primary and the total mass of the two eclipsing stars.


\begin{acknowledgements}
We are grateful to the support group of the 1.5-m, and 2.2-m-telescopes
at La Silla, especially Lisa Germany, Olivier Hainaut, Emanuela Pompei,
John Pritchard, Linda Schmidtobreick, Fernando Selman, Rolando Vega,
Erich Wenderoth, for helping us with the observations.  {We would also
like to thank the referee G. Cutispoto for sending us his spectra of BS
Ind and for helping us improve the manuscript}.  This research has
made use of the SIMBAD database, operated at CDS, Strasbourg, France.

\end{acknowledgements}

\end{document}